\RequirePackage{ifpdf}
\ifpdf % We are running pdfTeX in pdf mode
\documentclass[pdftex]{sigma}
\else
\documentclass{sigma}
\fi

\begin{document}

\allowdisplaybreaks

\renewcommand{\PaperNumber}{035}

\FirstPageHeading

\renewcommand{\thefootnote}{$\star$}

\ShortArticleName{Charges in Gauge Theories}

\ArticleName{Charges in Gauge Theories\footnote{This paper is a
contribution to the Proceedings of the O'Raifeartaigh Symposium on
Non-Perturbative and Symmetry Methods in Field Theory (June
22--24, 2006, Budapest, Hungary). The full collection is available
at \href{http://www.emis.de/journals/SIGMA/LOR2006.html}{http://www.emis.de/journals/SIGMA/LOR2006.html}}}

\Author{David MCMULLAN}
\AuthorNameForHeading{D. McMullan}

\Address{School of Mathematics and Statistics, University of Plymouth, Plymouth, UK}
\Email{\href{mailto:dmcmullan@plymouth.ac.uk}{dmcmullan@plymouth.ac.uk}}
\URLaddress{\url{http://www.tech.plym.ac.uk/maths/staff/dmcmullan/home.html}}

\ArticleDates{Received October 09, 2006, in f\/inal form February
09, 2007; Published online February 28, 2007}

\Abstract{Recent progress in the construction of both electric, coloured and magnetic charges in gauge theories will be presented. The topological properties of the charged sectors will be highlighted as well as the applications of this work to conf\/inement and infrared dynamics.}


\Classification{81U20; 81V05; 81V10}

\section{Introduction}
The particle data book~\cite{Yao:2006} contains a wealth of information on the particles that build up the standard model. We can f\/ind masses, magnetic moments, estimates and  bounds on their sizes, and much more besides. All of this f\/ills us with admiration for the skill and ingenuity of our experimental colleagues. This should be contrasted with our theoretical understanding of these particles where even the humble electron resists a mathematical description. Indeed, in their seminal paper~\cite{Kulish:1970ut} on QED, Kulish and Faddeev conclude with the striking statement that ``the relativistic concept of a charged particle does not exist''. The situation in non-Abelian theories is even murkier where the expectation is that, for example, colour forces conf\/ine coloured particles and yet there is no description of just what a coloured particle is meant to be!

The response of the theoretical community to this state of af\/fairs is surprisingly pragmatic:  experiments measure inclusive cross-sections and these we can calculate. There is no need to worry about a lack of a particle description for these processes as this is all hidden in the unavoidable f\/inite resolutions of the experiments. This would be f\/ine (although not necessarily attractive)  if we had a secure theoretical understanding of how to calculate such cross-sections. But we don't and are forced to restrict attention to observables that are insensitive to the infrared.

In this talk I want to summarise how we have arrived at the above situation and then discuss a programme of work, which has been developed in collaboration with many colleagues, that tries to do better. The successes to date that we have achieved will be presented and issues that we are still grappling with will be pointed out. This work started while I was a scholar in DIAS in 1991. In those early fragile  stages of this  project, when ideas were not fully formed and connections only hinted at, I valued the encouragement and advice given by Lochlainn who was always happy to question the party line.

\section{The infrared}
It is the masslessness of the photon that makes the large scale, infrared properties of QED so confusing. This was recognised in the early days of QED by Bloch and Nordsieck \cite{Bloch:1937pw} who argued that in any scattering process with f\/inal state electrons, there will inevitably be emission of soft (undetectable) photons. Summing over these leads to a infrared f\/inite, but f\/inal state resolution dependent, cross-section.

It is helpful to recall in a little more detail how this cancellation of the infrared divergences works. For example, in Coulombic scattering, the one-loop virtual contribution to the cross-section has an infrared divergence which is proportional to $-1/\epsilon$, where $\epsilon$ is the infrared regulator that arises by using  dimensional regularisation. Summing over the soft-emissions shown in Fig.~1 yields an additional infrared divergence now proportional to $1/\epsilon$. Combining these \emph{degenerate} processes cancels the infrared pole in the cross-section: $(-1+1)/\epsilon=0$.

\begin{figure}[h]
 \begin{center}
\parbox{30mm}{\includegraphics[width=25mm]{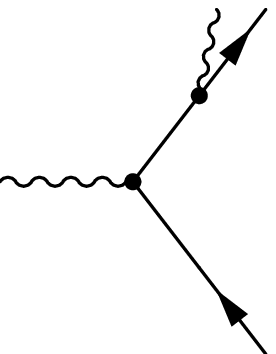}}
    + \
\parbox{30mm}{\includegraphics[width=25mm]{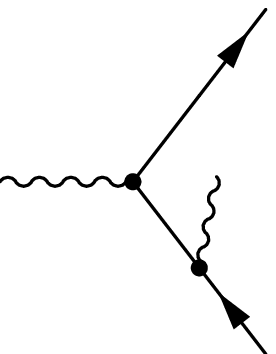}}
\end{center}
 \caption{Soft photon emission from the in-coming and out-going electron.}
 \end{figure}

The Bloch--Nordsieck approach has the great advantage of yielding f\/inite results, but is it the right thing to do? There is a strange asymmetry in it whereby out-going electrons are accompanied by soft photons but in-coming ones are not! It seems to assume that experimentalists have complete control over their in-states but are a bit vague on what is going out. To rephrase this: degeneracies are essential for the out-states but not allowed for the in-states. This might just seem a minor quibble were it not for the well known fact that Bloch--Nordsieck fails if the charges are also massless. To overcome this we have the celebrated Lee--Nauenberg theorem~\cite{Lee:1964is} which says that \emph{all} degeneracies, including now initial and f\/inal, must be summed over to get infrared f\/inite cross-sections.

Although physically appealing, this is a surprising claim as a naive extension of the Bloch--Nordsieck method to soft-absorption (as described in Fig.~2) would introduce a new set of infrared divergences that are identical to the ones that arose through soft-emission. The overall infrared pole thus becomes proportional to $(-1+1+1)/\epsilon\ne0$.
\begin{figure}[h]
\begin{center}
\parbox{30mm}{\includegraphics[width=25mm]{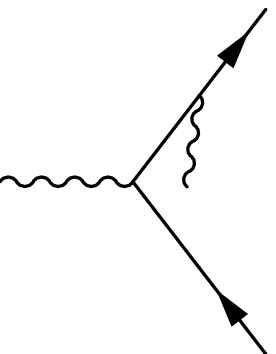}}
    + \
\parbox{30mm}{\includegraphics[width=25mm]{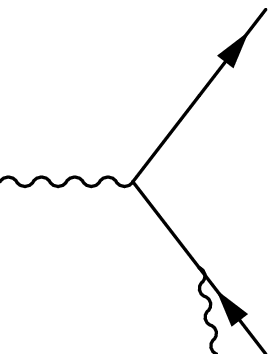}}
\end{center}
 \caption{Soft photon absorption from the in-coming and out-going electron.}
 \end{figure}

A more careful analysis of the Lee--Nauenberg method~\cite{Lavelle:2005bt} shows that allowing both initial and f\/inal state degeneracies leads to a host of new and unexpected processes that need to be carefully  summed over. For example, in Fig.~3 we have the processes whereby an initial state photon is absorbed and emitted by the electrons.
\begin{figure}[h]
\begin{center}
\parbox{30mm}{\includegraphics[width=25mm]{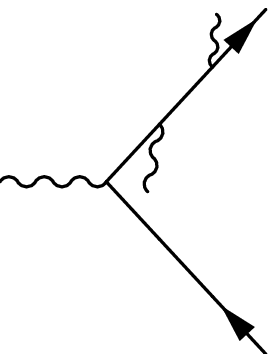}}
    + \
\parbox{30mm}{\includegraphics[width=25mm]{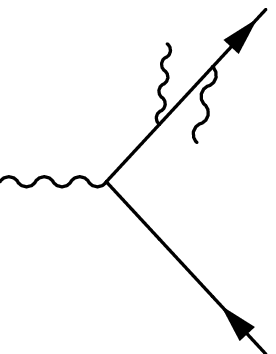}}
+ \
\parbox{30mm}{\includegraphics[width=25mm]{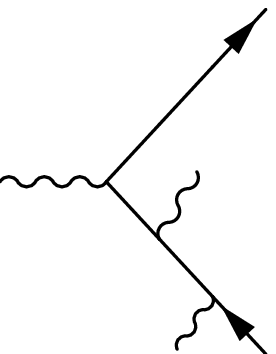}}\\[.3cm]
\parbox{30mm}{\includegraphics[width=25mm]{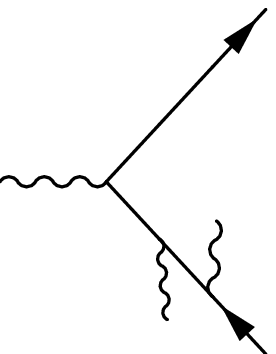}}
    + \
\parbox{30mm}{\includegraphics[width=25mm]{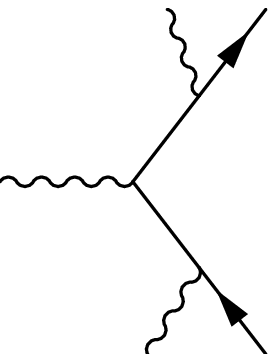}}
+ \
\parbox{30mm}{\includegraphics[width=25mm]{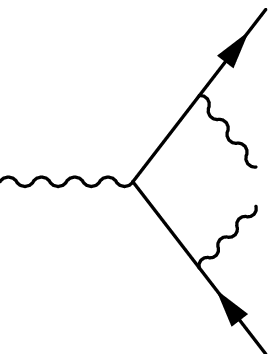}}
\end{center}
 \caption{Soft photon emission and absorption from the in-coming and out-going electron.}
 \end{figure}

These process contribute to the cross-section at this order only through their interference with the disconnected process described in Fig.~4.

 \begin{figure}[h]
\begin{center}
\parbox{30mm}{\includegraphics[width=25mm]{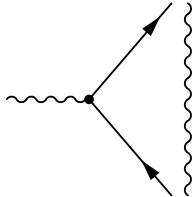}}
\end{center}
 \caption{A degenerate but disconnected process.}
 \end{figure}

Usually such disconnected processes are not allowed to contribute to the cross-section but here they are essential for the infrared cancellation implied by the Lee--Nauenberg theorem (see the full discussion of this in~\cite{Lavelle:2005bt}). However, even this is not the full story. Including these degenerate processes contributes, among other things, a $-2/\epsilon$ to the cross-section. So including these leads to the overall infrared pole: $(-1+1+1-2)/\epsilon\ne0$. There is still a residual infrared divergence!

To f\/inally arrive at an infrared f\/inite cross-section, where we have both initial and f\/inal state degeneracies, we can extend this discussion and simply include more disconnected  processes as these still contribute at the same order in perturbation theory. So, for example,  we can include the situation shown in Fig.~5, where we have emission and a~disconnected line. This gives a~connected contribution to the cross-section which is identical to the original emission process. Including this we f\/inally arrive at the infrared cancellation
\[
(-1+1+1-2+1)\frac{1}{\epsilon}=0.
\]

\begin{figure}[h]
\begin{center}
\parbox{30mm}{\includegraphics[width=25mm]{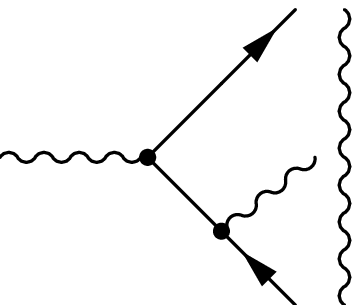}}
   \ + \
\parbox{30mm}{\includegraphics[width=25mm]{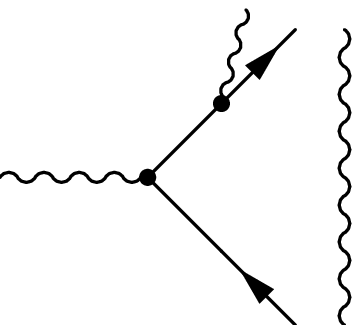}}
\end{center}
 \caption{A degenerate initial state photon and two f\/inal state photons.}
 \end{figure}

This seems to be what we were after: an infrared f\/inite cross-section including both initial and f\/inal state degeneracies. However, once this Pandora's box of disconnected processes is opened, we have to follow the logic through to the bitter end. Why stop with these processes? Surely we should sum over \emph{all} degenerate processes that contribute at this order in perturbation theory. In doing this we f\/ind~\cite{Lavelle:2005bt} that the infrared cancellation slips through our f\/ingers and we are left  with an ill-def\/ined theory.  This is not just a problem with soft divergences, but also arises when there are initial and f\/inal state massless charges which generate additional  collinear infrared divergences. Thus we see that the Lee--Nauenberg approach to infrared f\/inite cross-sections is not well def\/ined and hence one of the basic building blocks to our understanding of how charges scatter has been removed.

\section{Constructing charges}
In the 60's the infrared problem in QED  was revisited by several authors culminating in the work of Kulish and Faddeev~\cite{Kulish:1970ut} who showed that the masslessness of the photon implies that the interaction between electrons and photons never vanishes even at asymptotic times. As a~consequence the  propagator of the matter has a branch cut rather than a pole as you approach the mass-shell. Hence the loss of a particle description.

However, the fact that the interaction never ``switches of\/f'' means that the matter f\/ield never becomes gauge invariant and hence physical. The natural question to ask then is: can we f\/ind a physical (i.e., gauge invariant) f\/ield that does correspond asymptotically to a particle? To make this concrete, suppose that we want to identify the f\/ield that asymptotically corresponds to a~static charge, and for simplicity, let's initially only work in the Abelian theory.

We want to f\/ind a gauge invariant operator $\psi_0(x)$ which allows us to construct asymptotically a charged state from the vacuum via $\left|\psi_0(x)\right>=\psi_0(x)\left|0\right>$. This state should have the correct electric and magnetic f\/ields associated with it at asymptotic times:
\[
E_i(x)\left|\psi_0(r)\right>=\frac{{e}}{4\pi}
\frac{(x-r)_i}{|\underline{x}-\underline{r}|^3}\left|\psi_0(r)\right>,\qquad B_i(x)\left|\psi_0(r)\right>=0.
\]

The massiveness of the electron means that we can succinctly characterise the asymptotic region of soft interactions as that where the mass scale of the matter f\/ield dominates. In an ef\/fective description of such a static heavy f\/ield we can use the equation of motion
\[
D_0\psi=(\partial_0-ieA_0)\psi=0.
\]
Thus we can def\/ine a static f\/ield $\tilde{\psi}$ via
\[
\tilde{\psi}(x)=\exp\left(-ie\int_a^{x_0}A_0(s,\underline{x})\,ds\right)\psi(x),
\]
where we have suppressed an overall (anti-)time ordering. Using the equations of motion it is straightforward to show that $\partial_0\tilde{\psi}=0$.

This is not yet the f\/ield which generates a static charge since it is not gauge invariant. To make it physical while still maintaining its static properties, we perform a f\/ield dependent gauge transformation with parameter $T{\cdot} A/T{\cdot}\partial$ where $T$ is to be specif\/ied but has, at least, to be such that the operator $T{\cdot} \partial$ is invertible.

Under this transformation we have
\[
A_\mu\to A_\mu-\partial_\mu\left(\frac{T{\cdot} A}{T{\cdot}\partial}\right)
\]
and
\[
\psi\to\exp\left(-ie\frac{T{\cdot} A}{T{\cdot}\partial}\right)\psi.
\]
Both of these f\/ields are now gauge invariant and we see that on the transformed f\/ield we have $T{\cdot} A=0$.

Acting on our static conf\/iguration $\tilde{\psi}$ we get the gauge invariant result
\[
\psi_T(x)=\exp\left(-ie\int_a^{x_0}\frac{T^\mu F_{\mu0} }{T{\cdot}\partial}(s,\underline{x})\,ds\right)\exp\left(-ie\frac{T{\cdot} A}{T{\cdot}\partial}\right)\psi(x).
\]
We now need to specify the form for $T$ that is appropriate for a static charge. There are various ways~\cite{Usannals,Usannals1} to discuss this but perhaps the most direct way is to investigate the electric and magnetic f\/ields produced by this charge. In order to recover the expected Coulombic f\/ield discussed above we must take $T^0=0$ and $T^i=\partial^i$, in which case we arrive at the physical description of a static charge given by
\begin{equation}\label{charge}
    \psi_0(x)=\exp\left(ie\int_a^{x_0}\frac{\partial^i F_{i0} }{\nabla^2}(s,\underline{x})\,ds\right)\exp\left(-ie\frac{\partial_i A_i}{\nabla^2}\right)\psi(x)\,.
\end{equation}
This construction can be easily generalised to moving and, to some extent, to non-Abelian charges.

The f\/irst thing to note is that there is an interesting structure to this construction of charges. The f\/inal two terms in~(\ref{charge}) are, together, gauge invariant and represent the \emph{minimal} description of a charged f\/ield. Indeed, for the static charge, this form was written down by Dirac many years ago~\cite{Dirac:1955uv}. The \emph{additional} part of the construction is independently gauge invariant. This structure to the construction ref\/lects dif\/ferent physical aspects of the charge and becomes signif\/icant when studying the infrared properties~\cite{Usannals,Usannals1} or the potential between coloured charges~\cite{Bagan:2005qg}.

The extent to which this construction works in the non-Abelian theory has been investigated~\cite{Lavelle:1995ty} and reveals a fascinating interplay between the global aspects of the Yang--Mills conf\/iguration space and the ability to construct coloured charges. Relating this obstruction more directly to conf\/inement is a dif\/f\/icult but ongoing task~\cite{Ilderton:2007qy}.

\section{Magnetic charges}
Having seen the extent to which electric and coloured charges can be constructed, it is natural to ask whether a similar construction can be found for magnetic charges. This is particularly relevant to any investigation into the conf\/inement of coloured charges as it is common lore that the  condensation of magnetic
monopoles is responsible for conf\/inement. However, although there has been numerous lattice investigations of this, many questions have been left open. An analytic description is lacking in the literature to complement and support these lattice investigations. So what we want to investigate now is the extent to which we can construct a  gauge invariant description of a monopole operator.

A magnetic monopole operator $M(r)$ should clearly be gauge invariant and allow us to construct a one monopole state via its action on the vacuum:
\[
\left| M(r)\right>:=M(r)\left|0\right>.
\]
In addition, it should create a Coulombic magnetic f\/ield distribution
\[
B_i(x)\left|
M(r)\right>=\frac{1}{g}\frac{(x-r)_i}{|\underline{x}-\underline{r}|^3}\left|
M(r)\right>,
\]
and have some stability properties such as f\/inite energy, etc. Finding such an operator is far from straightforward.

Let us start f\/irst with the Abelian theory. We know from Dirac that monopoles can only arise if singular potentials are used. Standard examples are:
\[
\underline{\lambda}^{N}:=-\tfrac12g\frac{\underline{r}\times
\hat{\underline{z}}}{r(r+z)},\qquad\underline{\lambda}^{S}:=\tfrac12g\frac{\underline{r}\times
\hat{\underline{z}}}{r(r-z)},
 \]
which are, respectively, regular of\/f the neqative/positive $z$-axis.

Using these potentials we can immediately construct a  candidate operator which is manifestly gauge invariant
\[
M(r)=\exp\left(\frac{i}g\int d^3w \lambda^{N}_i(w-r)E_i(w)\right).
\]
However, in addition to producing the desired Coulombic magnetic f\/ield, this operator also produces the Dirac string which carries the return f\/lux and hence produces a conf\/iguration with no overall magnetic charge.

We can do better than this, though, if we mimic the bundle construction of monopoles by removing the
position of the monopole and   introduce a multi-valued
potential:
\[
\underline{\Lambda}(r)=\theta(z)\underline{\lambda}^{N}+\theta(-z)\underline{\lambda}^{S}
+\frac1g \delta(z)\phi(r)\underline{\hat{z}},
\]
where $\phi$ is the polar angle.
Armed with this we can construct an improved operator
\[
M(r)=\exp\left(\frac{i}g\int_{\mathbb{R}^3-\{r\}} d^3w
\Lambda_i(w-r)E_i(w)\right),
\]
which is still gauge invariant but now generates only the Coulombic magnetic f\/ields.

There is now a temptation to push ahead and extend this construction to QCD. However, that is a very dif\/f\/icult step and it is not at all obvious how monopoles should be described in a pure gauge theory. Instead we shall pursue a more conservative route and look to the Georgi--Glashow model which has bona f\/ide monopole conf\/igurations.

We recall that the Georgi--Glashow model is an SU(2) gauge f\/ield coupled to adjoint Higgs
  \[
L=-\tfrac14 F^2+(DH)^2-V(H^2).
  \]
In this theory we can def\/ine a \emph{gauge invariant} Abelian f\/ield strength
  \[
F_{\mu\nu}=\frac{H^a}{|H|}F^a_{\mu\nu}-\frac1g\frac1{|H|^3}\epsilon^{abc}H^a(D_\mu
H)^b(D_\nu H)^c.
\]
Hence we have a  magnetic current
\[J_\mu^M=\frac12\epsilon_{\mu\nu\lambda\sigma}\partial^{\nu}F^{\lambda\sigma}\]
and a magnetic charge exists as a physical observable.
\[
Q_M=\frac1{4\pi}\int d^3x J_0^M=\frac1{8\pi g}\int d^2 S_i
\epsilon_{ijk}\epsilon^{abc}\hat{H}^a\partial_j \hat{H}^b\partial_k
\hat{H}^c.
\]

In this theory we f\/ind~\cite{Khvedelidze:2005rv} that a candidate monopole operator can be constructed of the form
\[
M(r)=D(r)M_A(r),
\]
where we f\/irst create the monopole and its string
\[
M_A(r)=\exp\left(\frac{i}g\int d^3w
\lambda^{N}_i(w-r)\hat{H}^a(w)E^a_i(w)\right),
\]
then we remove string contribution to the magnetic f\/ield
\[
D(r)=\exp\left(\!\frac{i}g\int\!\!\! d^3w
\phi(r-w)\delta(r_{\perp}-w_{\perp})\frac{r_i-w_i}{|r_{\perp}-w_{\perp}|}
\hat{H}^a(w)E^a_i(w)\!\right).
\]
This is again multi-valued but is now allowed by the (vanishing) Higgs f\/ields.

To test this construction we recall that in the Higgs phase monopoles are massive so we expect that in a
 f\/inite but large volume $L$,
 \[
   \left<M\right>\propto\exp\left(-\mu L\right).
 \]
 In contrast to this, in the conf\/ining phase we expect that
  \[
\left<M\right>\ne0.
  \]
  This is, however, a non-perturbative ef\/fect. In a perturbative calculation, though, we'd expect a milder volume
  dependence and this we were able to test.
  We were able to show within a path integral calculation, using the steepest descent
  method and a specif\/ic conf\/iguration that we called the dandelion conf\/iguration, that
\[
\left<M\right>=\exp\left(-\frac{c}{g^2}\ln(\Lambda L)\right),
\]
which is the milder volume dependence expected.

\section{Conclusions}
The short distance, ultraviolet structure of gauge theories is
well understood. In contrast, the large scale, infrared properties
of gauge theories is full of intricacies and surprises that are
only gradually coming to light and are still far from being
understood. We have seen in this article that some progress can be
made in the dif\/f\/icult task of constructing  charges in both
Abelian and non-Abelian theories, and these have already given
some insight into the soft infrared divergences that arise in
scattering processes, and the structure of the inter-quark
potential. We have also seen the route to the construction of
magnetic charges. However, much remains to be done. In particular,
the following are immediate but still open questions:
\begin{itemize}\itemsep=0pt
  \item How can we construct well def\/ined cross-sections that are infrared f\/inite when we allow both initial and f\/inal state degeneracies? More specif\/ically, how do we deal with the disconnected contributions?
  \item How do we extend the description of  charges to massless ones? Will we still see structure in the construction and what would  this correspond to physically?
  \item How do we describe the impact of the global topological structures found in the Yang--Mills conf\/iguration space on the global description of coloured charges?
  \item How can we construct monopoles in a pure gauge theory?
\end{itemize}

\subsection*{Acknowledgements}
This work has grown out of many intense collaborations over the past years, most notably that with Martin Lavelle. I'd also like to acknowledge many conversations with Emili Bagan, Robin Horan, Arsen Khvedelidze and Alex Kovner.

\pdfbookmark[1]{References}{ref}
\LastPageEnding

\end{document}